\documentclass[useAMS,usenatbib]{mn2e}

\usepackage{epsfig}
\usepackage{amsmath}
\usepackage{amssymb}
\usepackage{natbib}
\usepackage{threeparttable} 
\usepackage{booktabs}

\usepackage{epstopdf}
\usepackage{times}

\newcommand{\chan}{\textit{Chandra}}
\newcommand{\swift}{\textit{Swift}}

\newcommand{\xmm}{\textit{XMM-Newton}}
\newcommand{\inte}{\textit{Integral}}

\newcommand{\rosat}{\textit{ROSAT}}

\newcommand{\nustar}{\textit{NuSTAR}}

\newcommand{\Msun}{\mathrm{M}_{\odot}}
\newcommand{\lum}{\mathrm{erg~s}^{-1}}
\newcommand{\flux}{\mathrm{erg~cm}^{-2}~\mathrm{s}^{-1}}

\newcommand{\cnts}{\mathrm{c~s}^{-1}}

\newcommand{\nh}{\mathrm{cm}^{-2}}
\newcommand{\gmc}{GM/c^2}
\newcommand{\dist}{(D/5.0~\mathrm{kpc})^2}

\newcommand{\ledd}{L_{\mathrm{Edd}}}
\newcommand{\lx}{L_{\mathrm{X}}}

\newcommand{\kms}{\mathrm{km~s}^{-1}}
\newcommand{\risco}{R_{\mathrm{ISCO}}}

\newcommand{\relx}{\textsc{relxill}}

\newcommand{\source}{IGR J17062--6143}

\def \nar {NewAR}

\def \mnras {MNRAS}
\def \apj {ApJ}
\def \apjs {ApJS}
\def \apjl {ApJL}
\def \aap {A\&A}
\def \nat {Nature}

\def \atel {ATel}

\def \ssr {SSRv}

\def \physrep {Phys. Rev.}
\def \apss {A\&AS}

\title[Accretion in \source]{An in-depth study of a neutron star accreting at low Eddington rate: On the possibility of a truncated disk and an outflow}
\author[N. Degenaar et al.]
{N. Degenaar$^{1,2}$\thanks{e-mail: degenaar@ast.cam.ac.uk}, C. Pinto$^1$, J.M.~Miller$^{3}$, R.~Wijnands$^2$, D. Altamirano$^{4}$, F. Paerels$^{5,6}$, 
\newauthor 
A.C. Fabian$^{1}$ and D.~Chakrabarty$^{7}$\\ 
$^1$Institute of Astronomy, University of Cambridge, Madingley Road, Cambridge CB3 OHA, UK\\
$^2$Anton Pannekoek Institute for Astronomy, University of Amsterdam, Science Park 904, 1098 XH, Amsterdam, the Netherlands\\
$^3$Department of Astronomy, University of Michigan, 1085 South University Avenue, Ann Arbor, MI  48109, USA\\
$^4$Department of Physics and Astronomy, University of Southampton, Southampton, Hampshire, SO171BJ, UK\\
$^5$Columbia University, Mail Code 5246, 550 West 120th Street, New York, NY 10027, USA\\
$^6$Columbia Astrophysics Laboratory, Mail Code 5247, 550 West 120th Street, New York, NY 10027, USA\\
$^7$Massachusetts Institute of Technology (MIT), Kavli Institute for Astrophysics and Space Research, Cambridge, MA 02139, USA
}

\begin{document}

\date{Accepted 2016 September 12. Received 2016 September 12; in original form 2016 August 16}

\pagerange{\pageref{firstpage}--\pageref{lastpage}} \pubyear{0000}

\maketitle

\label{firstpage}

\begin{abstract}
Due to observational challenges our knowledge of low-level accretion flows around neutron stars is limited. We present \nustar, \swift\ and \chan\ observations of the low-mass X-ray binary \source, which has been persistently accreting at $\simeq$0.1 per cent of the Eddington limit since 2006. Our simultaneous \nustar/\swift\ observations show that the 0.5--79 keV spectrum can be described by a combination of a power law with a photon index of $\Gamma$$\simeq$2, a black body with a temperature of $kT_{\mathrm{bb}}$$\simeq$$0.5$~keV (presumably arising from the neutron star surface), and disk reflection. Modeling the reflection spectrum suggests that the inner accretion disk was located at $R_{\mathrm{in}}$$\gtrsim$$100~\gmc$ ($\gtrsim$225~km) from the neutron star. The apparent truncation may be due to evaporation of the inner disk into a radiatively-inefficient accretion flow, or due to the pressure of the neutron star magnetic field. Our \chan\ gratings data reveal possible narrow emission lines near 1~keV that can be modeled as reflection or collisionally-ionized gas, and possible low-energy absorption features that could point to the presence of an outflow. We consider a scenario in which this neutron star has been able to sustain its low accretion rate through magnetic inhibition of the accretion flow, which gives some constraints on its magnetic field strength and spin period. In this configuration, \source\ could exhibit a strong radio jet as well as a (propeller-driven) wind-like outflow.
\end{abstract}

\begin{keywords}
accretion: accretion disks -- pulsars: general -- stars: individual (\source) -- stars: neutron -- X-rays: binaries -- X-rays: bursts
\end{keywords}

%%%%%%%%%%%%%%%%%
% INTRODUCTION
%%%%%%%%%%%%%%%%%

\section{Introduction}\label{sec:introduction}
Low-mass X-ray binaries (LMXBs) contain a neutron star or a black hole that accretes gas from a less massive companion star. These are excellent laboratories to study accretion in the strong gravity regime. In particular, LMXBs are observed over a wide range in X-ray luminosity, hence accretion rate, allowing the investigation of different accretion morphologies.

LMXBs are most easily discovered and studied when their X-ray luminosity is a sizable fraction of the Eddington limit, $\lx$$\gtrsim$0.1$~\ledd$. However, a lot of accretion activity occurs at much lower $\lx$. Indeed, many LMXBs are transient and accrete at $\lx$$>$$0.1~\ledd$ for only a few weeks or months and then spend years in quiescence before a new outburst commences. Thermal-viscous instabilities in the accretion disk provide the general framework for understanding such outburst-quiescence cycles \citep[though it cannot explain details; e.g.][for a review]{lasota01}. 

Accretion does not necessarily switch off in quiescence and may persist down to very low $\lx$ \citep[e.g.][]{wagner1994,campana1997,rutledge2002_aqlX1,kuulkers2008,cackett2010_cenx4,cackett2013_cenx4,bernardini2013,chakrabarty2014_cenx4,dangelo2014,rana2016}. Furthermore, there is a growing population of LMXBs that exhibit outbursts with a peak luminosity of only $\lx$$\simeq$$10^{-4}-10^{-2}~\ledd$ \citep[e.g.][]{hands04,sakano05,muno05_apj622,wijnands06,degenaar09_gc,campana09,heinke2010,armas2011,armas2014,sidoli2011}.  Understanding the properties of LMXBs at $\lx$$<$$10^{-2}~\ledd$ is thus an important part of forming a complete picture of accretion flows around neutron stars and stellar-mass black holes.

At $\lx$$>$$10^{-2}~\ledd$, matter is typically transferred through a geometrically thin and optically thick accretion disk that extends close to the compact primary. When moving towards lower $\lx$, the inner disk is expected to evaporate into a hot, geometrically thick and radiatively inefficient accretion flow \citep[e.g.][]{narayan1994,blandford99,menou2000,dubus2001_dim}. The X-ray spectral softening observed as black hole LMXBs transition from outburst to quiescence seems to support the formation of a radiatively-inefficient accretion flow \citep[see e.g.][for discussions]{plotkin2013,reynolds2014_qbh,yang2015}.

Neutron star LMXBs also show X-ray spectral softening at $\lx$$<$$10^{-2}~\ledd$ \citep[e.g.][]{armas2011,armas2013_2,degenaar2013_xtej1709,bahramian2014,allen2015,weng2015}, although the behavior is not the same as for black holes \citep[][]{wijnands2015}. The X-ray spectra are also different; high-quality data obtained for neutron stars at $\lx$$\simeq$$10^{-4}~\ledd$ have revealed the presence of a thermal component, likely from the accretion-heated neutron star surface, and a power-law spectral component that is harder than for black holes at similar $\lx$ \citep[e.g.][]{armas2013,armas2013_2,degenaar2013_xtej1709,wijnands2015}. The picture proposed for neutron star LMXBs is that the radiation from the accretion flow softens at $\lx$$<$$10^{-2}~\ledd$ like in black holes, but becomes overwhelmed by the emission released when matter impacts the neutron star surface (causing thermal emission and a hard tail) at $\lx$$<$$10^{-3}~\ledd$ \citep[][]{wijnands2015}. Deep observations of the nearby neutron star Cen X-4 at $\lx$$\simeq$$ 4\times10^{-6}~\ledd$ may support this idea; it appears that only thermal emission from the stellar surface and bremsstrahlung from a boundary layer (where the accretion flow meets the surface) are observed, whereas the accretion flow itself is not directly detected \citep[][]{chakrabarty2014_cenx4,dangelo2014}. 

The magnetic field of a neutron star may also have a discernible effect on the accretion flow. It can potentially truncate the inner  disk and re-direct plasma along the magnetic field lines. X-ray pulsations from the heated magnetic poles may then be seen \citep[e.g.][]{pringle1972,rappaport1977,finger1996,wijnands1998}. X-ray spectral observations revealed truncated inner disks in several X-ray pulsars \citep[e.g.][]{miller2011,papitto2013_hete,degenaar2014_groj1744,king2016,pintore2016}, whereas the inner disk seems to extend further in for non-pulsating neutron star LMXBs \citep[e.g.][]{cackett2010_iron,miller2013_serx1,degenaar2015_4u1608,disalvo2015,ludlam2016}. Magnetic field effects can possibly gain importance when the accretion rate drops, allowing the magnetic pressure to increasingly compete with that exerted by the disk.

Since transient LMXBs typically only spend a short time at $\lx$$\simeq$$10^{-4}$--$10^{-2}~\ledd$, this is a particularly challenging accretion regime to characterize and to capture with sensitive observations. Fortuitously, a handful of neutron star LMXBs accrete in this range for several years \citep[e.g.][]{chelovekov07_ascabron,delsanto07,jonker08,heinke09_vfxt,zand09_J1718,degenaar2010_burst,degenaar2011_asca,armas2013}. These very-faint X-ray binaries (VFXBs) are interesting targets to further our knowledge of low-level accretion flows. 

VFXBs are also intriguing because the disk instability model has trouble explaining how their low accretion rates can be sustained for many years \citep[e.g.][]{dubus99,lasota01}. One possibility is that these objects have small binary orbits that only fit small accretion disks \citep[e.g.][]{king_wijn06,zand07,hameury2016}. The very dim optical counterparts indeed suggests that some VFXBs may have short orbital periods \citep[e.g.][]{bassa08,zand09_J1718}. However, a few other VFXBs were found to harbor H-rich donors, which rules out very compact orbits \citep[e.g.][]{degenaar2010_burst,arnason2015}. An alternative explanation for the quasi-stable low accretion rate of VFXBs is that the neutron star's magnetic field inhibits the accretion flow \citep[e.g.][]{wijnands2008_MC,heinke09_vfxt,heinke2014_gc,patruno2010_vf,degenaar2014_xmmsource}. 

%%%%

\subsection{The very-faint X-ray binary \source}
\source\ was discovered with \inte\ in 2006 \citep[][]{churazov2007}, but it was not until 2012 that it was identified as a neutron star LMXB through the detection of an energetic thermonuclear X-ray burst \citep[][]{degenaar2012_igrburster}. The X-ray burst light curve showed wild intensity variations that were presumably caused by extreme expansion of the neutron star photosphere; this suggests a source distance of $D$$\simeq$5~kpc \citep[][]{degenaar2013_igrj1706}, assuming that the peak flux of the X-ray burst reached the empirical Eddington limit of $L_{\mathrm{Edd}}$$=$$3.8\times10^{38}~\lum$ \citep[][]{kuulkers2003}. The source has been persistently accreting at a low luminosity of $L_X$$\simeq$$4 \times 10^{35}~\dist~\lum$ for the past 10 yr, which roughly corresponds to $\simeq$$10^{-3} \ledd$ \citep[e.g.][]{ricci2008,remillard2008,degenaar2012_igrburster}. It was not detected by the \rosat/PSPC in 1990 (obsID rs932824n00), suggesting that its luminosity was likely a factor $\gtrsim$10 lower at that time.

Due to its relative proximity and relatively low interstellar extinction compared to other VFXBs, \source\ is a particularly good target to further our understanding of low-level accretion flows and the nature of these peculiar LMXBs. In particular, the powerful X-ray burst seen in 2012 revealed the presence of Fe in the accreted matter \citep[][]{degenaar2013_igrj1706}, which provides the chance of detecting reflection features from the accretion disk. 

Disk reflection manifests itself most prominently as an Fe-K emission line at $\simeq$6.4--6.97~keV and a Compton hump at $\simeq$20--40 keV \citep[e.g.][]{george1991,matt1991}. The shape of these features is modified by Doppler and gravitational redshift effects as the gas in the disk moves in high-velocity Keplerian orbits inside the gravitational well of the compact accretor. The reflection spectrum thus encodes information about the accretion morphology \citep[e.g.][for a review]{fabian2010}. In particular, detecting and modeling disk reflection features allows for a measure of the inner radial extent of the accretion disk, $R_{\mathrm{in}}$. If a radiatively-inefficient accretion flow forms or if the stellar magnetic field is dynamically important in governing the accretion flow in VFXBs, the inner disk is expected to be truncated away from the neutron star. However, so far no observational constraints on the inner radial extent of the accretion disks in VFXBs have been obtained yet.

Radiatively-inefficient accretion flows are likely associated with outflows \cite[e.g.][]{rees1982,narayan1994,blandford99,narayan2005}. Furthermore, the magnetic field of a neutron star may act as a propeller and could expel (some of) the in-falling gas \citep[e.g.][]{illarionov1975,lovelace1999,romanova2009,papitto2015_prop}. Neutron star LMXBs accreting at $\lx$$<$$10^{-2}~\ledd$ may therefore be expected to exhibit outflows. In the X-ray band, these may reveal itself through the detection of blue-shifted narrow spectral lines. 

In this work we present \nustar, \swift\ and \chan/HETG observations of \source\ to study the accretion regime of $\lx$$\simeq$$10^{-3} \ledd$ and to understand the puzzling nature of neutron star LMXBs that are able to accrete at such a low rate for years. In particular, the aim of these observations was to constrain the continuum spectral shape, to measure disk reflection features to gain insight into the accretion geometry, and to search for narrow X-ray spectral lines that may be indicative of an outflow.

%%%%%%%%%%%%%%%%%
% OBSERVATIONS
%%%%%%%%%%%%%%%%%

\section{Observations and data analysis}

\subsection{\nustar}\label{subsec:nustar}
Our \nustar\ observation was performed between 19:26 \textsc{ut} on 2015 May 6 and 05:01 \textsc{ut} on May 8 (obs ID 30101034002). The two co-aligned focal plane modules (FPM) A and B provide an energy coverage of 3--79~keV. Standard processing with \textsc{nustardas} (v. 1.4.1), resulted in $\simeq$70~ks of on-target exposure time per module. We extracted light curves, spectra, and response files with \textsc{nuproducts}, using a circular region of $30''$ radius for the source and a void circular region of $60''$ radius on the same chip for the background. \source\ was detected at a constant intensity of $\simeq$$3~\cnts$ during the observation (3--79~keV, FPMA+B summed).

When fitting the FPMA/B spectra simultaneously with a constant factor floating in between, we found that the flux calibration agreed to within 0.5\%. We therefore opted to combine the spectra of the two mirrors using \textsc{addascaspec}. A weighted response file was created using \textsc{addrmf}. The combined spectrum was grouped into bins with a minimum of 20 photons using \textsc{grppha}. In the combined spectrum the source was detected above the background in the entire \nustar\ bandpass; the signal to noise ratio (SNR) was $\simeq$500 around 7~keV and $\simeq$2 around 70~keV.

\subsection{\swift/XRT}\label{subsec:swift}
\swift\ observed \source\ simultaneous with \nustar\ on 2015 May 6 from 23:29--23:44 \textsc{ut} (obs ID 37808005). A $\simeq$0.9~ks exposure was obtained during a single orbit and the XRT was operated in photon counting mode. The source was detected at a constant count rate of $\simeq$$0.9~\cnts$. Source and background spectra were extracted using \textsc{XSelect}. To circumvent the effects of pile-up, we used an annular region with inner--outer radii of $12''$--$71''$ for the source. Background events were extracted from a void region with an area three times larger than that used for the source. An arf was created with \textsc{xrtmkarf}, using the exposuremap as input. The appropriate rmf (v. 15) was taken from the \textsc{caldb}. The spectral data were grouped into bins with a minimum of 20 photons.

\subsection{\chan/HETG}\label{subsec:chan}
We observed \source\ with \chan\ on 2014 October 25 from 04:48 to 13:39 \textsc{ut} for $\simeq$29~ks of on-source exposure (obs ID 15749), and for $\simeq$64~ks from 09:34 \textsc{ut} on October 27 to 14:57 \textsc{ut} on October 28 (obs ID 17543). The High Energy Transmission Gratings (HETG) were used to disperse the incoming light on to the ACIS-S CCDs that were operated in faint, timed mode. The HETG consists of the Medium Energy Grating (MEG; 0.4--5~keV, 31--2.5~\AA) and the High Energy Grating (HEG; 0.8--8~keV, 15--1.2~\AA). We used the data from both grating arms.

The data were reprocessed using the \textsc{chan$\_$repro} script. We extracted light curves for the first orders of the HEG and MEG instruments at 100-s resolution using the task \textsc{dmextract}. \source\ was detected at similar, constant count rates of $\simeq$$1.9~\cnts$ during the two observations (MEG+HEG combined; 0.4--8~keV). We extracted the first order HEG and MEG spectra from the reprocessed pha2 files employing \textsc{dmtype2split}, and generated response files using \textsc{mktgresp}. 

We used \textsc{combine$\_$grating$\_$spectra} to combine the spectra of the plus and minus orders of each instrument, and to sum the data from the two observations. To investigate the spectral continuum shape and flux, we grouped the combined spectra to a minimum of 20 photons~bin$^{-1}$. To search for narrow spectral features, we instead fitted the un-binned spectra of the plus and minus orders separately. A constant multiplication factor was always included in the modeling to allow for calibration differences.  

\subsection{Spectral analysis}
For the analysis of the continuum and reflection spectrum, we used \textsc{xspec} \citep[v. 2.9][]{xspec} and applied $\chi^2$-statistics. Interstellar absorption was included in all our spectral fits using the \textsc{tbabs} model, employing cross-sections from \citet{verner1996} and abundances from \citet{wilms2000}. For our high-resolution spectral analysis of the \chan\ data, we instead employed \textsc{spex} (v. 3.0)\footnote{https://www.sron.nl/spex} because it is more suitable for finding and modeling narrow spectral features than \textsc{xspec}. Throughout this work we assume a source distance of $D$$=$5~kpc \citep[][]{degenaar2013_igrj1706}, an Eddington luminosity of $L_{\mathrm{Edd}}$$=$$3.8\times10^{38}~\lum$ \citep[][]{kuulkers2003}, and a neutron star mass and radius of $M$$=$1.4~$\Msun$ and $R$$=$10~km, respectively. All errors are given at 1$\sigma$ confidence level.

 \begin{figure}
 \begin{center}
		\includegraphics[width=8.5cm]{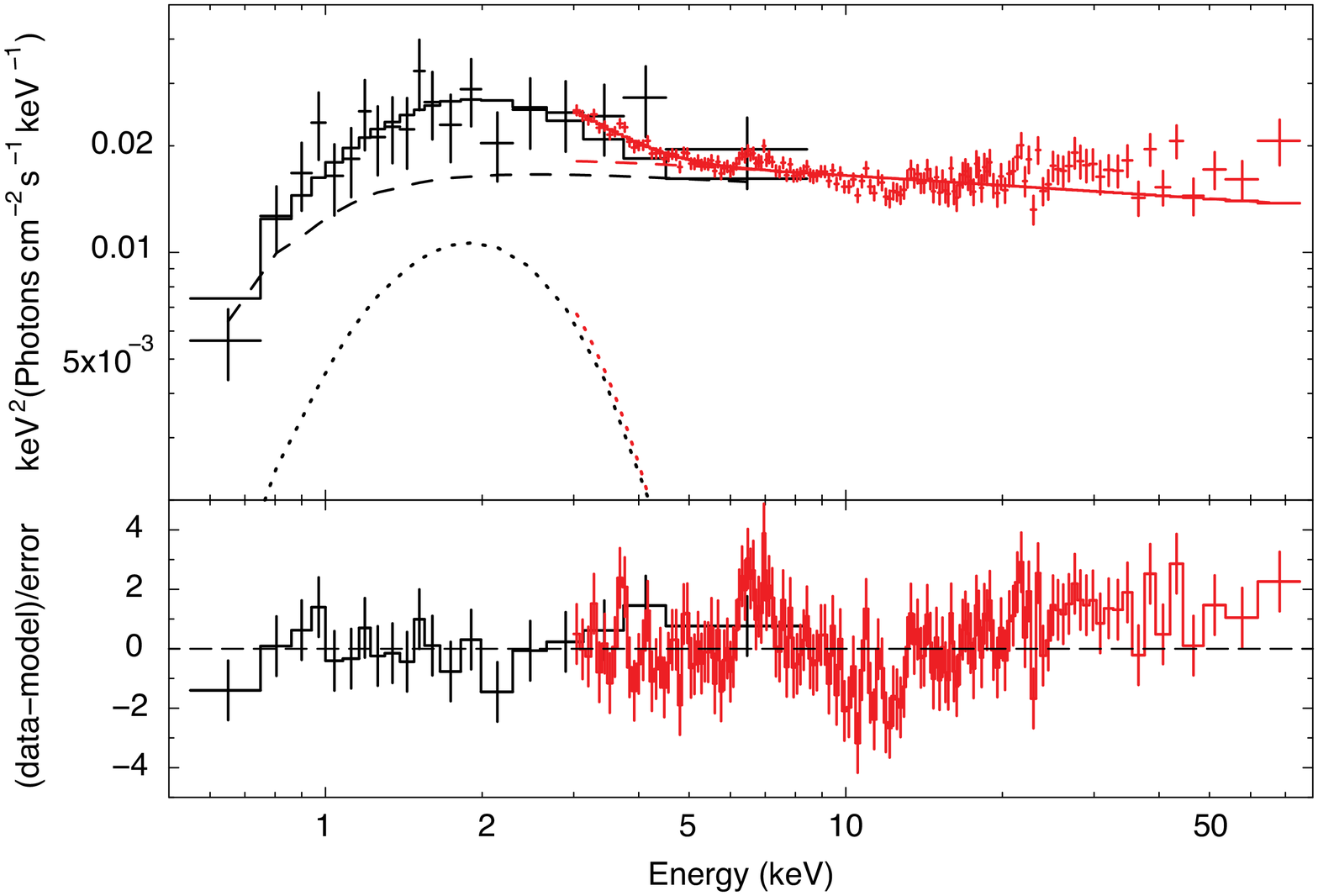}	
                  \includegraphics[width=8.5cm]{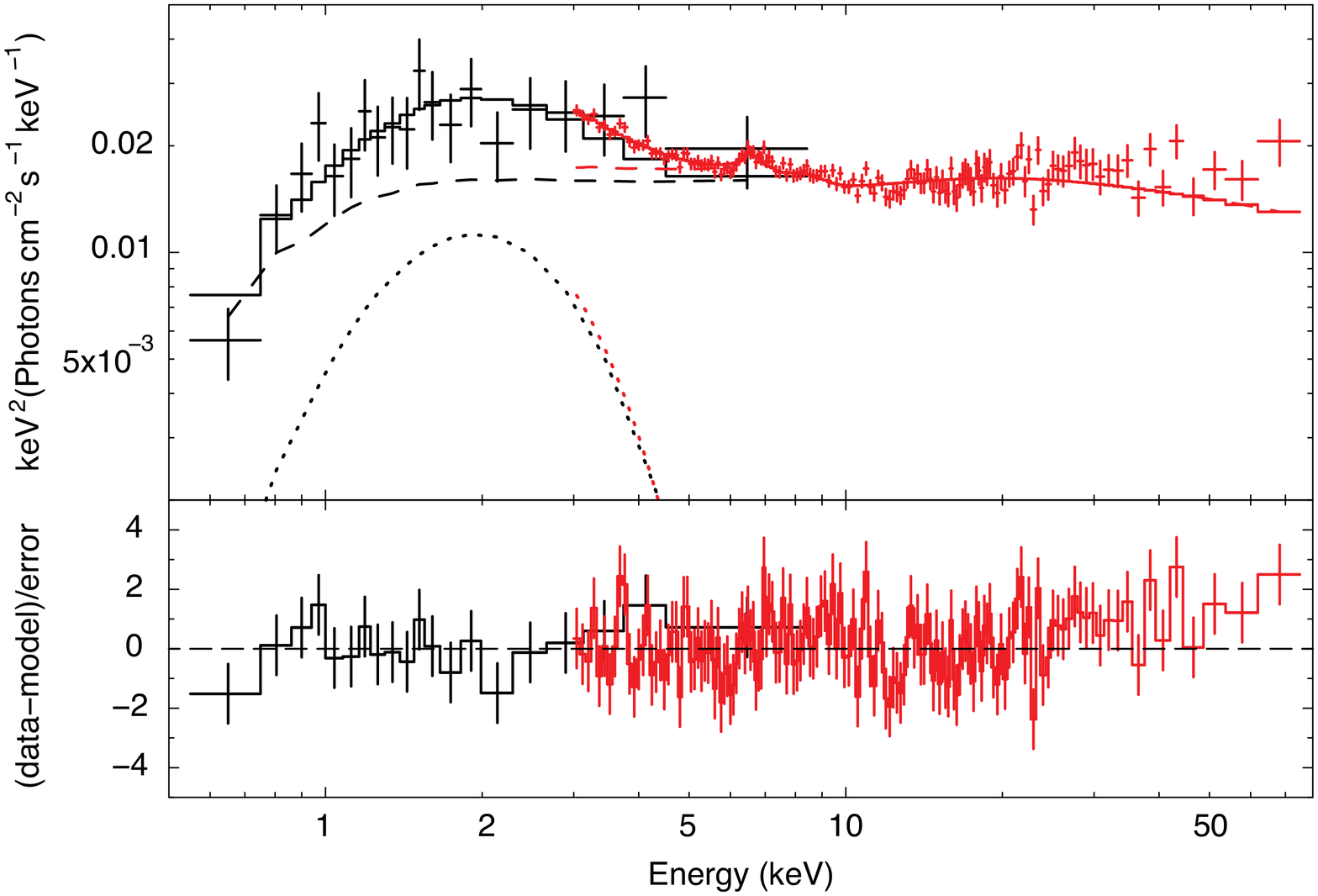}     
		\includegraphics[width=8.5cm]{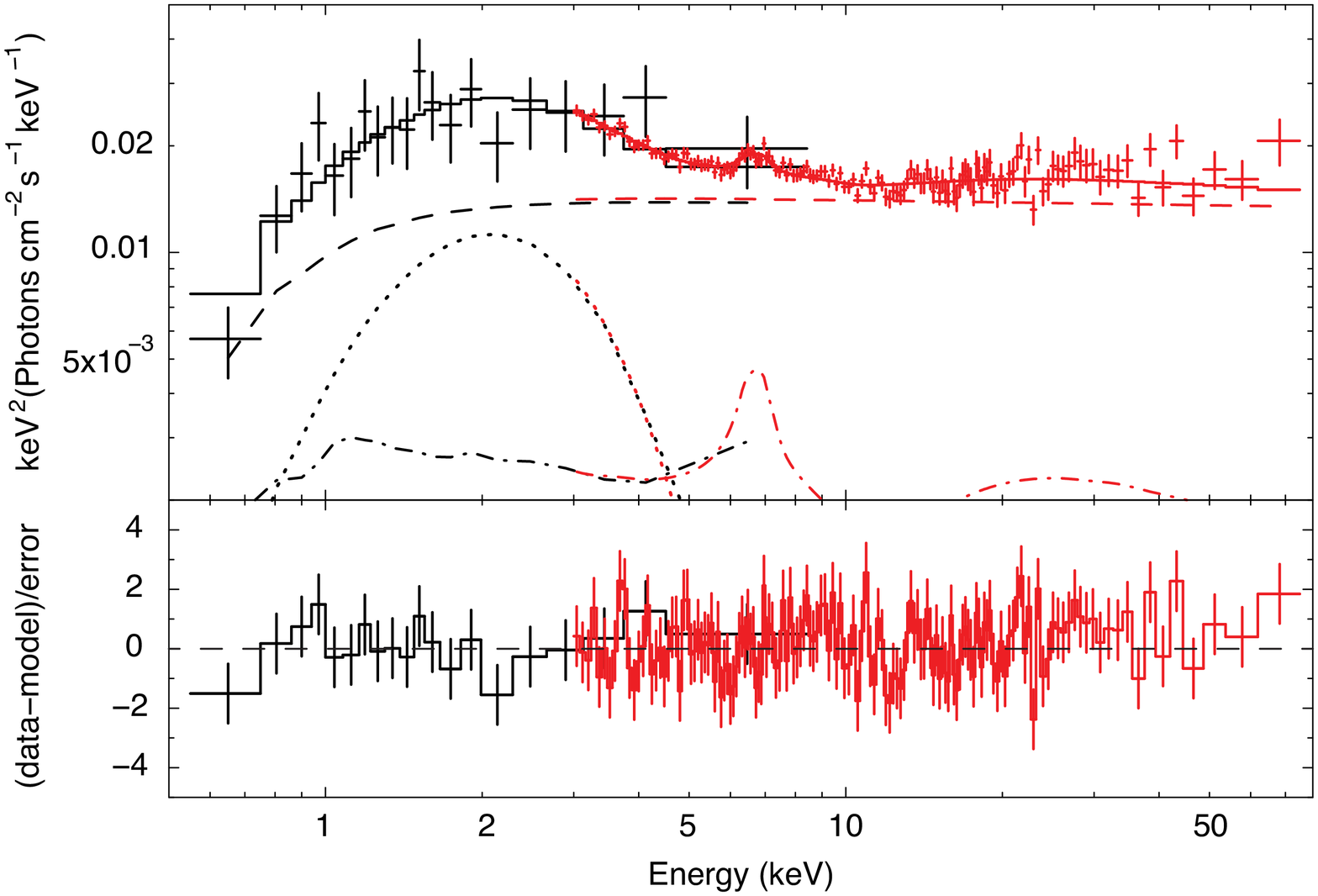}        
    \end{center}
    \caption[]{Fits to the \nustar\ (red) and \swift\ (black) spectral data. 
    The lower panels show the residuals in sigmas. Top: Continuum fit to an absorbed black body (dotted curve) and power law (dashed curve) model. Middle: Fit with the power-law component replaced by \textsc{relxill} to include relativistic reflection (dashed curve). Bottom: Fit with a black body (dotted curve) plus power law (dashed curve) continuum and \textsc{reflionx} (dashed-dotted curve) as the relativistic reflection model. We note that the highest \nustar\ and lowest \swift\ energy bins have a low SNR.
        }
 \label{fig:spec}
\end{figure}

%%%%%%%%%%%%%%%%%
% SPECTRA
%%%%%%%%%%%%%%%%%

\section{Results}

\subsection{\nustar/\swift\ continuum X-ray spectrum}\label{subsec:nucont}
A single power-law model (\textsc{pegpwrlw}) does not provide a good fit to the 0.5--79 keV \nustar/\swift\ spectrum ($\chi_{\nu}^2$$=$$ 1.52$ for 583 d.o.f.). Adding a soft, thermal component (\textsc{bbodyrad}) yields a significant improvement ($\chi_{\nu}^2$$=$$ 1.15$ for 581 d.o.f.; $F$-test probability $\simeq$$10^{-36}$). We obtain $N_{\mathrm{H}}$$=$$(2.3 \pm 0.4)\times10^{21}~\nh$ \citep[comparable to that found in previous studies; e.g.][]{ricci2008,degenaar2013_igrj1706}, $\Gamma$$=$$2.10\pm 0.01$, $kT_{\mathrm{bb}}$$=$$0.46\pm0.03$~keV and $R_{\mathrm{bb}}$$=$$3.8\pm0.4~\mathrm{(km/5.0~kpc)}^2$. Replacing the power-law component by a Comptonized emission component (\textsc{nthcomp}) yielded a worse fit ($\chi_{\nu}^2$$=$$ 1.21$, 580 d.o.f.). We also tried cutoff and broken power-law models, but these revealed no evidence for a detectable change in the power-law index in the 0.5--79 keV spectrum. 

Our \textsc{pegpwrlw+bbodyrad} continuum fit is shown in Figure~\ref{fig:spec} (top). The inferred 0.5--79 keV unabsorbed flux is $F_{0.5-79}$$=$$(1.17\pm0.02)\times10^{-10}~\flux$, which corresponds to $L_{0.5-79}$$=$$(3.50\pm0.06) \times 10^{35}~\dist~\lum$. This suggests that \source\ was accreting at $\simeq$$10^{-3} \ledd$ during our 2015 \nustar/\swift\ observations. The spectrum in the \nustar\ band is dominated by the power-law spectral component, which contributes $\simeq$97\% to the total unabsorbed 3--79 keV flux. For reference, the 0.5--10 keV luminosity inferred from this fit is $L_{0.5-10}$$\simeq$$1.6 \times 10^{35}~\dist~\lum$. In this energy band the power-law spectral component also dominates, accounting for $\simeq$79\% of the total unabsorbed flux.

\begin{figure}
 \begin{center}
	\includegraphics[width=8.5cm]{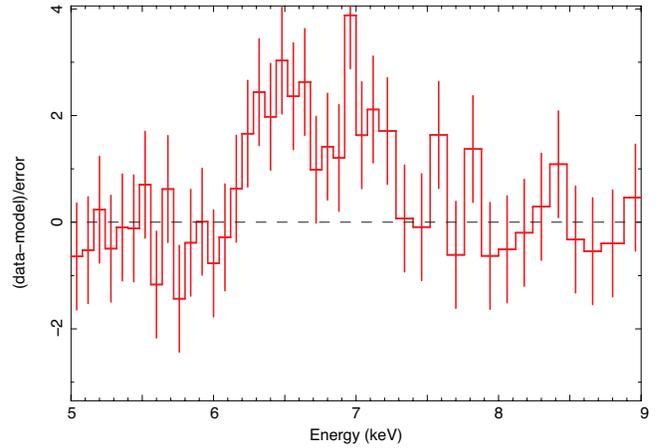}
    \end{center}
    \caption[]{Zoom of the \nustar\ data to model ratio in the Fe-K line region for an absorbed power law plus black body continuum model.
        }
 \label{fig:feline}
\end{figure}

Our continuum description leaves positive residuals near $\simeq$6--7~keV, as shown in Figure~\ref{fig:feline}. The feature is reminiscent of the broad Fe-K lines that are often seen in the X-ray spectra of bright LMXBs and are typically attributed to disk reflection.\footnote{See e.g. \citet{ng2010} for a discussion on alternative explanations, but see \citet{chiang2015} for a strong test case that favors disk reflection.} This motivates the inclusion of a relativistic reflection model in our spectral fits.

%%%%

\subsection{\nustar/\swift\ X-ray reflection spectrum}\label{subsec:refl}

\subsubsection{Reflection fits with \relx}\label{subsubsec:relx}
The shape of the reflection spectrum depends on the properties of the flux incident on the accretion disk. In case of \source, the X-ray spectrum is dominated by a $\Gamma$$\simeq$2 power-law component (Figure~\ref{fig:spec} top). Out of all the reflection models appropriate for an incident power-law spectrum, we chose \relx\ \citep[v. 0.4a;][]{garcia2014}, because it features higher spectral resolution and updated atomic data compared to other models. Furthermore, the reflection spectrum is calculated for each emission angle, which should be more accurate than averaging over different angles. The model combines the reflection grid \textsc{xillver} \citep[][]{garcia2013} with the convolution kernel \textsc{relconv} to include relativistic effects on the shape of the reflection spectrum \citep[][]{dauser2010}. Since it describes both the illuminating power-law and the reflected emission, we replaced \textsc{pegpwrlw} by \relx\ in our spectral fits. 

We set the model up using an unbroken emissivity profile of the form $\epsilon(r)$$\propto$$r^{-q}$. Reflection fits for LMXBs are typically not sensitive to the outer disk radius because the emissivity profile drops off steeply with increasing radius, so we fixed $R_{\mathrm{out}}$$=$$500~\risco$, where $\risco$ is the location of the innermost stable circular orbit (ISCO). The remaining model parameters are then the index $\Gamma$ and the high-energy cutoff $E_{\mathrm{cut}}$ of the illuminating power law, the dimensionless spin parameter $a$, the disk inclination $i$, the inner disc radius $R_{\mathrm{in}}$ (expressed in terms of $\risco$), the ionization parameter $\log \xi$, the iron abundance $A_{\mathrm{Fe}}$ (with respect to Solar), the reflection fraction $R_{\mathrm{refl}}$, and the normalization $N_{\mathrm{refl}}$. We note that figure~1 of \citet{fabian1989} gives an instructive overview of the effect of several of these parameters on the shape of relativistically-broadened Fe-K lines.

The spin of the compact object plays a role in setting $\risco$, but for neutron stars this is sufficiently low ($a$$\simeq$0--0.3) to be only a small effect \citep[e.g.][]{miller2013_serx1}. The spin of \source\ is unknown, but neutron stars in LMXBs are expected to be spun up to millisecond periods due to the angular momentum gained by accretion \citep[e.g.][]{alpar1982,bhattacharya1991,strohmayer1996,wijnands1998}. This is indeed borne out by observations; spin periods of $\simeq$1.6--10~ms have been inferred for about two dozen neutron star LMXBs from detecting coherent X-ray pulsations or rapid intensity oscillations during thermonuclear X-ray bursts \citep[e.g.][for a list]{patruno2010}. Here, we assumed $a$$=$0.3 based on the approximation $a$$\simeq$$0.47/P_{\mathrm{ms}}$, where $P_{\mathrm{ms}}$ is the spin period in ms \citep[valid for $M$$=$$1.4~\Msun$;][]{braje2000}. According to equation 3 from \citet{miller1998}, the ISCO is then located at $\risco$$\simeq$$6 [1-0.54a] \gmc$$\simeq$$5.05~\gmc$, which corresponds to $\simeq$11.2~km for $M$$=$$1.4~\Msun$. Indeed, setting $a$$=$0 (i.e. the Schwarzschild metric with $\risco$$=$$6~\gmc$) did not strongly affect our inferred inner disk radius (see below).

Several model parameters were not constrained by the data when left free to vary and were therefore fixed to avoid degeneracy. Firstly, we set $q$$=$3, which is theoretically motivated \citep[e.g.][]{wilkins2012} and often applies well to neutron star LMXBs \citep[e.g.][for a sample study]{cackett2010_iron}. Secondly, $E_{\mathrm{cut}}$ pegged at the model upper limit of 1 MeV. Since our continuum modeling showed no evidence for an observable cutoff in the \nustar\ passband we fixed $E_{\mathrm{cut}}$$=$$500$~keV. This is a somewhat arbitrarily high value but consistent with that found from \nustar\ reflection modeling of the neutron star LMXB 4U 1608--52 at $\lx$$\simeq$$10^{-2}~\ledd$ \citep[][]{degenaar2015_4u1608}. We later tested how different values of $E_{\mathrm{cut}}$ affected our fits (see below). We also found the disk inclination to be unconstrained, pegging at the upper limit $90\degr$. Such a high inclination is ruled out by the fact that \source\ does not show dips or eclipses in its X-ray emission, which suggests $i$$\lesssim$$75\degr$ \citep[e.g.][]{frank2002}. As there is no typical value for the inclination of neutron star LMXBs,
we therefore explored fits with $i$$=$$25\degr,45\degr$, and $65\degr$. Finally, we fixed $A_{\mathrm{Fe}}$$=$1 because it was poorly constrained when left free to vary. The only free fit parameters for \relx\ thus were $R_{\mathrm{in}}$, $\Gamma$, $\log \xi$, $R_{\mathrm{refl}}$, and $N_{\mathrm{refl}}$. 

Our spectral analysis results are summarized in Table~\ref{tab:spec}. Replacing the power law by \textsc{relxill} provides a significant improvement. The statistical quality of the fit with $i$$=$$65\degr$ ($\chi_{\nu}^2$$=$$ 1.01$ for 578 d.o.f.) is better than the fits with $i$$=$$25\degr$ and $45\degr$ ($\chi_{\nu}^2$$=$$ 1.02$ for 578 d.o.f.) at a $\simeq$2--2.5$\sigma$ level ($\Delta \chi^2$$=$4.31--6.34). The effect of increasing the inclination is to shift the blue wing of the Fe-K line to higher energy \citep[see e.g. figure 1 of][]{fabian1989}. The blue wing of the broad line in \source\ extends all the way up to $\simeq$7.3~keV (Figure~\ref{fig:feline}), which may be reason that a higher inclination is preferred in our reflection fits. We opted to proceed our analysis with $i$$=$$65\degr$, but note that our conclusions are not affected by this choice. Our baseline fit described above is shown in Figure~\ref{fig:spec} (middle). We note that the highest \nustar\ and lowest \swift\ energy bins deviate from the model fit. This is plausibly due to a low SNR, but we explore the possible presence of an additional emission component in Section~\ref{subsec:powcomp}.

For $i$$=$$65\degr$, we obtained $R_{\mathrm{in}}$$=$$44.4^{*}_{-24.6}~\risco$ with the upper bound hitting the model limit of $100~\risco$. This inferred inner disk radius corresponds to $\simeq$$224^{*}_{-125}~\gmc$ for our choice of $a$$=$0.3, or $\simeq$$497^{*}_{-278}$~km for a neutron star mass of $M$$=$$1.4~\Msun$. Notably, the different inclinations yield comparable 1$\sigma$ lower limits of $R_{\mathrm{in}}$$\gtrsim$$20~\risco$ ($\gtrsim$$100~\gmc$), suggesting a truncated disk regardless of the chosen $i$. However, we note that the $\chi^2$-space is rather flat and $R_{\mathrm{in}}$ is still consistent with a location at the ISCO at $\simeq$3$\sigma$ confidence for our fits with $i$$=$65$\degr$. 

We explored the effect of our model assumptions for the obtained value of $R_{\mathrm{in}}$, our main parameter of interest. Table~\ref{tab:diffpar} summarizes the effect of changing the values of fixed model parameters ($a$, $q$, $E_{\mathrm{cut}}$, $A_{\mathrm{Fe}}$, and $R_{\mathrm{out}}$). This shows that the 1$\sigma$ lower limit on $R_{\mathrm{in}}$ always lies near $\simeq$$20~\risco$. A high cutoff energy appears to be preferred by our fits, which is consistent with the lack of an observable high-energy roll-over in the \nustar\ data. The reflection fits also appear to favor a high Fe abundance; this enhances the strength of the Fe-K line \citep[e.g.][]{ross2005}.

\begin{table}
\caption{Results from modeling the \nustar/\swift\ reflection spectrum.}
\begin{threeparttable}
\begin{tabular*}{0.48\textwidth}{@{\extracolsep{\fill}}lccc}
\hline
Model parameter  & \multicolumn{3}{c}{ \relx  } \\
\hline
 & $i=25\degr$ & $i=45\degr$ & $i=65\degr$   \\
$C$ (XRT)     & $0.92 \pm 0.05$ & $0.92 \pm 0.08$ & $0.92 \pm 0.08$  \\
$N_{\mathrm{H}}$ ($\times10^{21}~\nh$)  & $2.43 \pm 0.26$ & $2.47 \pm 0.35$ & $2.42 \pm 0.36$  \\
%&  &  &  \\
$kT_{\mathrm{bb}}$ (keV)   & $0.48 \pm 0.02$ & $0.47 \pm 0.02$  & $0.48 \pm 0.02$  \\
$N_{\mathrm{bb}}$ (km/10~kpc)$^2$   & $52.1^{+2.0}_{-12.0}$ & $53.5^{+20.5}_{-12.4}$ & $50.8^{+9.9}_{-13.5}$  \\
%&  &  &  \\
$\Gamma$   & $2.05\pm 0.01$ & $2.05 \pm 0.02$ & $2.05 \pm 0.02$  \\
$R_{\mathrm{in}}$ ($\risco$)   & $97.6^{*}_{-61.3}$ & $98.1^{*}_{-78.9}$ & $44.4^{*}_{-24.6}$   \\
$\log \xi$    & $3.28 \pm 0.04$ & $3.27 \pm 0.08$ & $3.19 \pm 0.11$  \\
$R_{\mathrm{refl}}$   & $0.24 \pm 0.03$  & $0.47 \pm 0.05$  & $0.48 \pm 0.09$   \\
$N_{\mathrm{refl}}$ ($\times 10^{-4}$)  & $2.41\pm 0.45$  & $2.37 \pm 0.10$  & $2.29 \pm 0.11$   \\
%&  &  &  \\
$\chi_{\nu}^2$ (dof)  & 1.02 (578) & 1.02 (578) & 1.01 (578)  \\
\hline
\end{tabular*}
\label{tab:spec}
\begin{tablenotes}
\item[] Notes. -- The constant factor $C$ was set to 1 for \nustar\ and left free for the \swift\ data. An asterisk indicates that a parameter pegged at the model limit.
Quoted errors reflect 1$\sigma$ confidence levels.
\end{tablenotes}
\end{threeparttable}
\end{table}

\begin{table}
\caption{Effect of fixed \relx\ parameters on the inner disk radius.}
\begin{threeparttable}
\begin{tabular*}{0.48\textwidth}{@{\extracolsep{\fill}}lcc}
\hline
Parameter value  & $R_{\mathrm{in}}$ ($\risco$) & $\Delta \chi^2$ \\
\hline
$a=0$ & $36.6^{*}_{-16.3}$ & $0$ \\
$q=2.5$ & $41.8^{*}_{-24.8}$ & $0.01$ \\
$q=4$ & $46.7^{*}_{-23.6}$ & $-0.01$ \\
$E_{\mathrm{cut}}=250$ & $49.0^{*}_{-16.8}$ & 3.4 \\
$E_{\mathrm{cut}}=750$ & $49.5^{*}_{-29.1}$ & $-2.5$ \\
$A_{\mathrm{Fe}}=0.5$ & $48.0^{*}_{-27.8}$ & 22.1 \\
$A_{\mathrm{Fe}}=2.5$ & $38.4^{*}_{-17.7}$ & $-16.8$ \\ 
$R_{\mathrm{out}}=1000~\risco$ & $37.0^{*}_{-17.2}$ & 0.1 \\ 
\hline
\end{tabular*}
\label{tab:diffpar}
\begin{tablenotes}
\item[] Notes. -- The $\Delta \chi^2$ value is given with respect to our baseline model with $i$$=$$65\degr$ (see Table~\ref{tab:spec}). An asterisk indicates that $R_{\mathrm{in}}$ pegged at the model limit of $100~\risco$.
Quoted errors reflect 1$\sigma$ confidence levels.
\end{tablenotes}
\end{threeparttable}
\end{table}

%%%%

\subsubsection{Reflection fits with \textsc{reflionx}}\label{subsubsec:otherrefl}
To ascertain that our inferred inner disk radius is not biased by our choice of model, we also fitted the reflection spectrum with \textsc{reflionx} \citep[][]{ross2005}. This model was convolved with \textsc{relconv} to allow relativistic effects to shape the reflection spectrum. The \textsc{reflionx} model also assumes that the illuminating flux is supplied by power-law spectrum, which is a good approximation for \source\ (see Section~\ref{subsec:nucont}). Setting $i$$=$65\degr, $q$$=$3, and $A_{\mathrm{Fe}}$$=$$1$ yielded a good fit ($\chi_{\nu}^2= 0.99$ for 578 d.o.f.) with $R_{\mathrm{in}}$$=$$53.2^{*}_{-20.0}~\risco$.  This fit using \textsc{reflionx} is shown in Figure~\ref{fig:spec} (bottom). Modeling the reflection spectrum with \textsc{reflionx} also favors a truncated inner disk.

%%%%

\subsection{Looking for a second power-law emission component}~\label{subsec:powcomp}
The presence of a $\simeq$0.5~keV black body in the spectrum of \source\ suggest that we likely observe radiation from the accretion-heated neutron star surface. As noted by \citet{wijnands2015}, this thermal emission may be associated with a hard emission tail that has an equal contribution to the flux in the 0.5-10 keV band as the thermal component. The hard tail should roll-over due to the spectral cutoff of the surface emission \citep[][]{dangelo2014}. We might thus expect the presence of a hard power-law emission component in the spectrum of \source, in addition to the observed thermal surface emission and the $\Gamma$$\simeq$2 power law that is presumably emitted by the accretion flow. 

Adding a second power law to our \textsc{bbodyrad+pegpwrl} model for the \nustar/\swift\ continuum gives a significant improvement ($\chi_{\nu}^2$$= $$1.09$ for 579 d.o.f.; $F$-test probability $\simeq$$10^{-7}$). The first power-law component has an index of $\Gamma_{1}$$=$$2.28 \pm 0.09$ and contributes $\simeq$77 per cent to the unabsorbed flux in the 0.5--10 keV band. The second power-law component is both fainter and harder: $\Gamma_{2}$$=$$1.11 \pm 0.37$ and a $\simeq$2 per cent contribution to the 0.5--10 keV unabsorbed flux. The fractional contribution of the hard power law to the 0.5--10 keV flux is considerably lower than that of the thermal emission ($\simeq$21\%), contrary to that expected when both are due to accretion on to the neutron star surface. Moreover, the second, harder power-law component starts to dominate the spectrum only at energies $\gtrsim$50~keV. It is plausible, that the harder power-law component is trying to account for spectral residuals at higher energies (see Figure~\ref{fig:spec} top) that could also be due to un-modelled reflection (i.e. a Compton hump). 

We also tried adding a second power-law component to a model that includes reflection (\relx\ with $i$$=$65\degr). The power-law folded into  \relx\ then has an index of $\Gamma_1$$=$$2.20\pm 0.03$. The additional power-law component is harder with $\Gamma_2$$=$$1.25^{+0.11}_{-0.50}$ and improves the fit ($\chi_{\nu}^2$$= $$0.99$ for 576 d.o.f.). However, an $F$-test suggests a $\simeq$$4\times10^{-4}$ probability ($\simeq$3.5$\sigma$) that this improvement is due to chance. Moreover, this additional power-law component starts to dominate the X-ray spectrum only at $\gtrsim$65~keV, which is close to the upper end of the \nustar\ energy band where the SNR is low (Section~\ref{subsec:nustar}). It is therefore not clear if an additional emission component, approximated by a power law, is indeed present in the broad-band spectrum of \source. We note that the inclusion of a second power-law component in the reflection fits leaves the inner disk radius virtually unchanged ($R_{\mathrm{in}}$$=$$41.5^{*}_{-19.6}~\risco$).

%%%%

\subsection{\chan\ X-ray continuum and reflection spectrum}\label{subsec:chancont}
\begin{figure}
 \begin{center}
	\includegraphics[width=8.5cm]{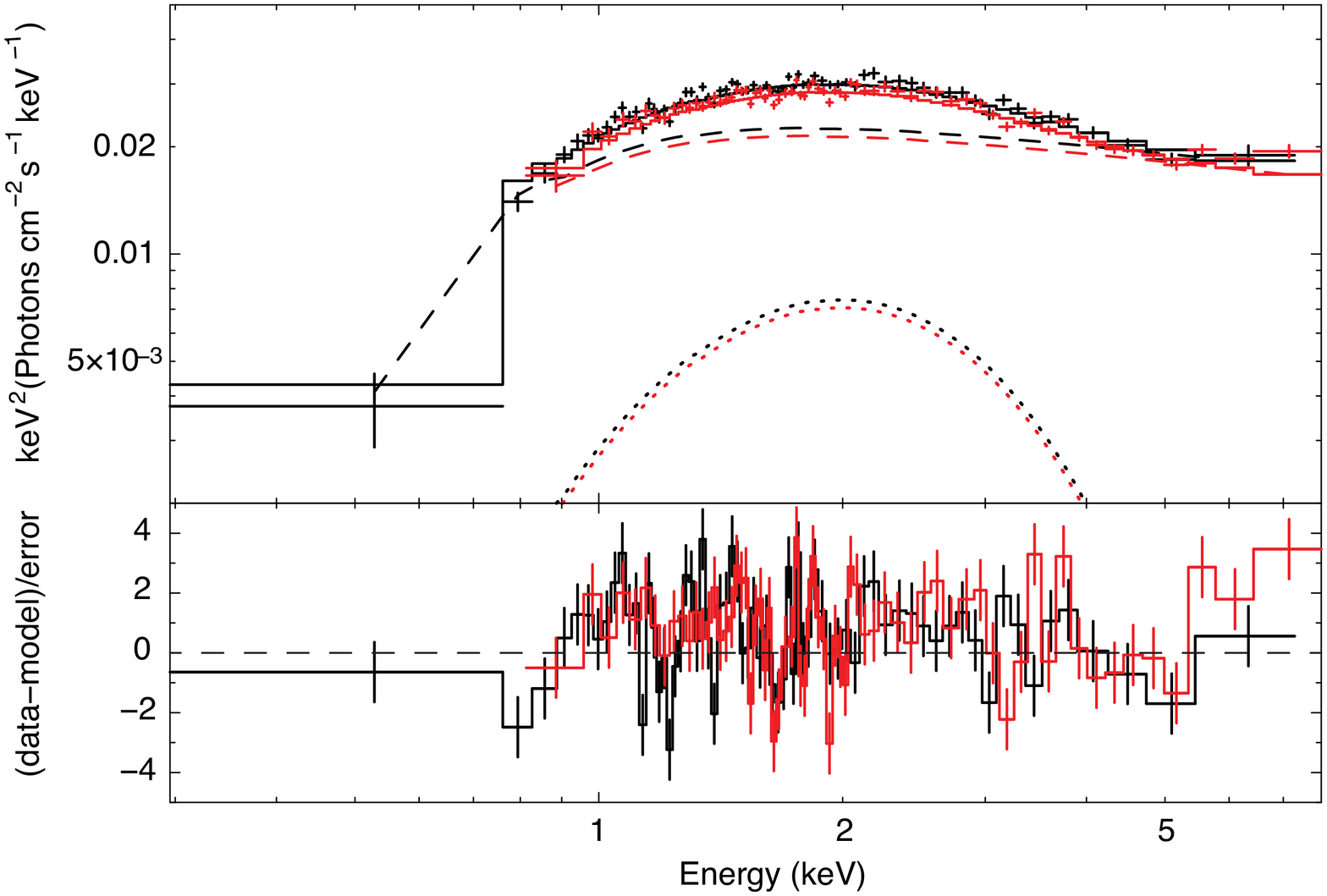}
	\includegraphics[width=8.5cm]{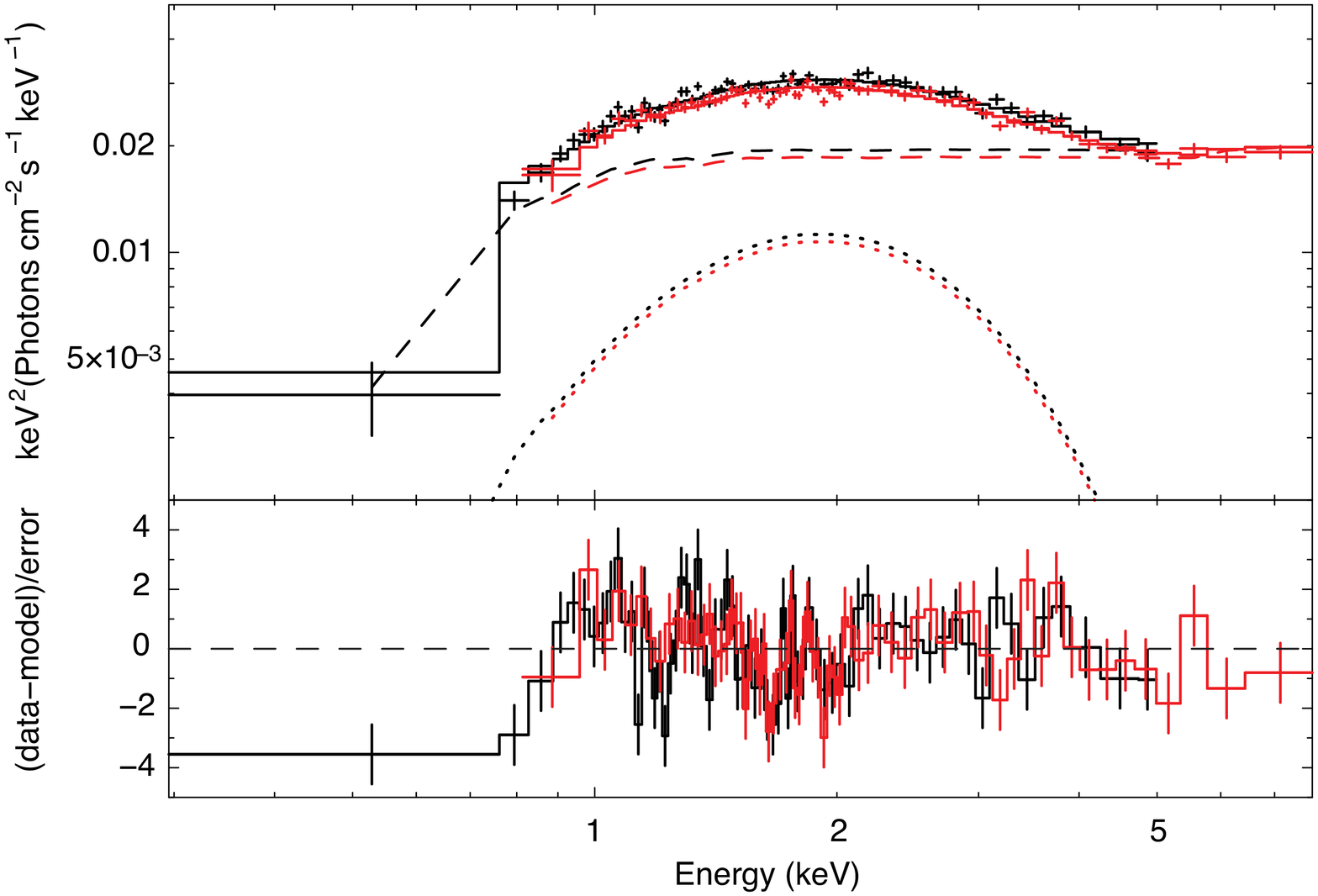}	
	\includegraphics[width=8.5cm]{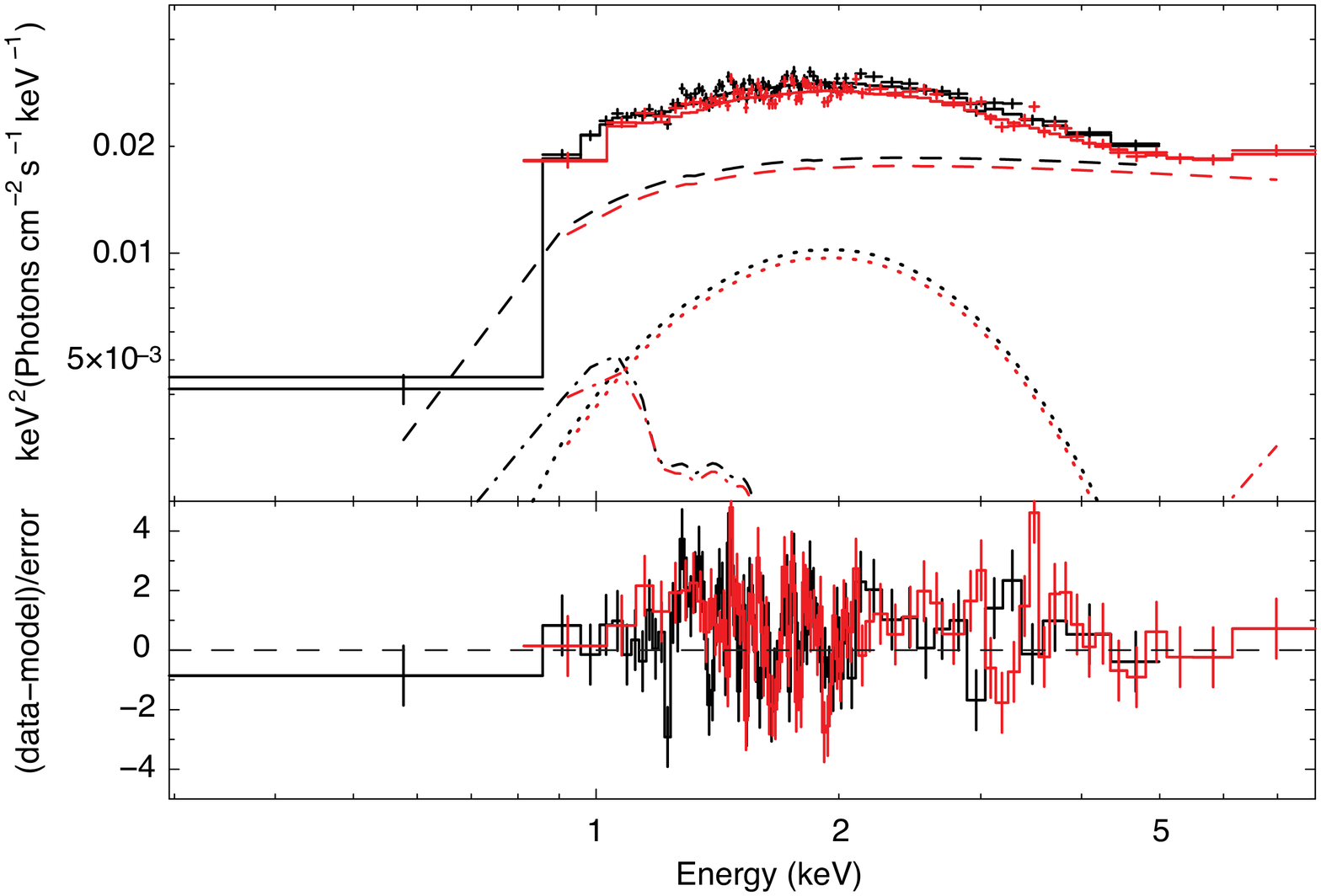}		
    \end{center}
    \caption[]{Fits to the \chan\ MEG (black) and HEG (red) spectral data. 
    Lower panels show the residuals in sigmas. Top: Continuum fit to an absorbed black body (dotted curve) and power law (dashed curve) model. Middle: Fit with the power-law component replaced by \relx\ to include relativistic reflection (dashed curve). Bottom: Fit with a black body (dotted curve) plus power law (dashed curve) continuum and \textsc{reflionx} (dashed-dotted curve) as the relativistic reflection model.
        }
 \label{fig:specchan}
\end{figure}

Just as for the \nustar/\swift\ data, the \chan\ spectrum is better described by a power law plus black body model ($\chi_{\nu}^2$$=$1.53 for 3955 d.o.f.) than by a power law alone ($\chi_{\nu}^2$$=$$1.56$ for 3957 d.o.f.; $F$-test probability $\simeq$$10^{-17}$). We obtain $N_{\mathrm{H}}$$=$$(2.6 \pm 0.3)\times10^{21}~\nh$, $\Gamma$$=$$2.24\pm 0.05$, $kT_{\mathrm{bb}}$$=$$0.48\pm0.01$~keV, and $R_{\mathrm{bb}}$$=$$2.77\pm 0.06~\mathrm{(km/5.0~kpc)}^2$. This continuum fit is shown in Figure~\ref{fig:specchan} (top). We measure a 0.5--10 keV unabsorbed flux of $F_{0.5-10}$$\simeq$$1.1\times10^{-10}~\flux$. This corresponds to $L_{0.5-10}$$\simeq$$3.2 \times 10^{35}~\dist~\lum$ and is a factor $\simeq$2 higher than observed in 2015 with \nustar/\swift\ (Section~\ref{subsec:nucont}).

\begin{table}
\caption{Results from modeling the \nustar/\chan\ reflection spectrum.}
\begin{threeparttable}
\begin{tabular*}{0.48\textwidth}{@{\extracolsep{\fill}}lccc}
\hline
Model parameter & \relx & \relx & \textsc{reflionx}  \\
\hline
 & $A_{\mathrm{Fe}}=1$ & $A_{\mathrm{Fe}}$ free  &  $A_{\mathrm{Fe}}$ free \\
$C$ (MEG)     & $1.09 \pm 0.01$ & $1.09 \pm 0.01$    & $1.09 \pm 0.01$   \\
$C$ (HEG)     & $1.02 \pm 0.01$ & $1.02 \pm 0.01$   & $1.02 \pm 0.01$   \\
$N_{\mathrm{H}}$ ($\times10^{21}~\nh$)  & $1.90 \pm 0.19$ & $1.68 \pm 0.11$   &  $2.47 \pm 0.03$   \\
%&  &    &     \\
$kT_{\mathrm{bb}}$ (keV)   & $0.50 \pm 0.01$ & $0.50 \pm 0.01$   &  $0.48 \pm 0.01$  \\
$N_{\mathrm{bb}}$ (km/10~kpc)$^2$   & $30.8 \pm 1.8$ & $35.4 \pm 1.2$   & $39.3 \pm 2.8$   \\
%&   &   &    \\
$\Gamma$   & $2.09 \pm 0.01$ & $2.04 \pm 0.01$   &  $2.13 \pm 0.02$  \\
$R_{\mathrm{in}}$ ($\risco$)   & $20.4 \pm 4.4$ & $36.0 \pm 9.5$   &  $36.0 \pm 9.5$  \\
$\log \xi$    & $3.04 \pm 0.05$ & $3.10 \pm 0.06$   & $2.56 \pm 0.10$   \\
$A_{\mathrm{Fe}}$ ($\times$~Solar)   & 1.0 fix & $5.0 \pm 0.5$   & $1.8 \pm 0.4$  \\
$R_{\mathrm{refl}}$   & $0.47 \pm 0.05$ & $0.26 \pm 0.03$   &  -  \\
$N_{\mathrm{refl}}$ ($\times 10^{-4}$) & $2.52 \pm 0.10$ & $2.64 \pm 0.06$    & $0.002 \pm 0.001$  \\
%&   &    &   \\
$\chi_{\nu}^2$ (dof)  & 1.47 (4516) & 1.46 (4515)    &  1.46 (4515)  \\
\hline
\end{tabular*}
\label{tab:specchan}
\begin{tablenotes}
\item[] Notes. -- The constant factor $C$ was set to 1 for \nustar\ and left free for the \chan\ data. Quoted errors are 1$\sigma$ confidence. 
\end{tablenotes}
\end{threeparttable}
\end{table}

 \begin{figure}
 \begin{center}
	\includegraphics[width=8.5cm]{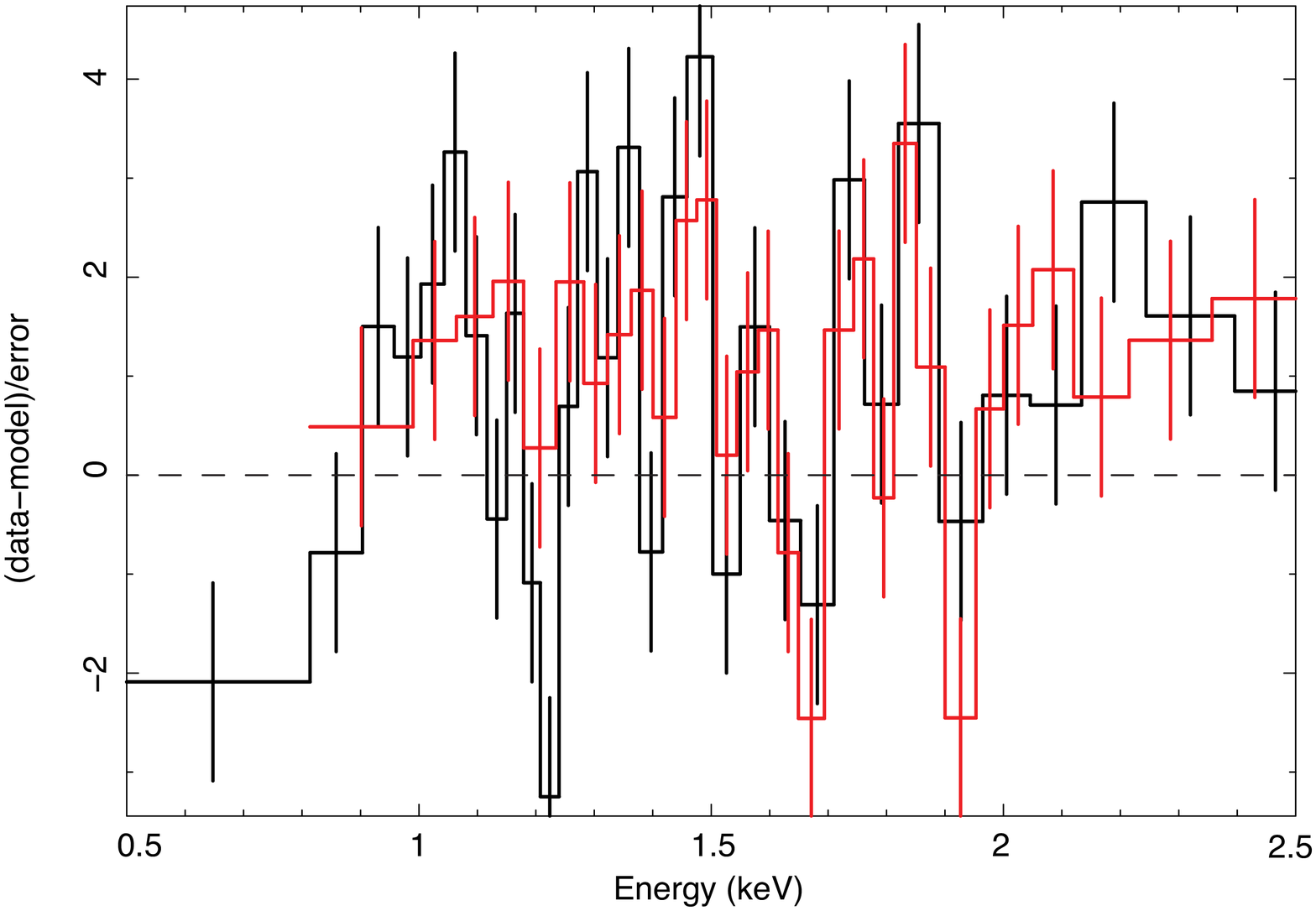}
    \end{center}
    \caption[]{Zoom of the \chan\ MEG (black) and HEG (red) data to model ratio for an absorbed power law plus black body continuum model.
        }
 \label{fig:lowElines}
\end{figure}

Our continuum fit leaves residuals at energies of $\simeq$1--2 and $\gtrsim$5~keV (Figure~\ref{fig:specchan}, top). A zoom of the 0.5--2.5 keV range is shown in Figure~\ref{fig:lowElines}. The un-modeled structure above 5~keV likely corresponds to a broadened Fe-K line, as was seen in the \nustar\ data and successfully modeled as disk reflection. The structure near 1~keV can potentially also arise from disk reflection \citep[e.g.][]{fabian2009}. 
To explore this possibility, we fitted the MEG and HEG data together with the \nustar\ spectrum. 

We used \textsc{relxill} plus a black body with all parameters tied between the different instruments, but with constant factor included to allow for flux and calibration differences. Similarly to our \nustar/\swift\ reflection fits, we fixed $q$$=$3, $R_{\mathrm{out}}$$=$$500~\risco$, $a$$=$0.3, $E_{\mathrm{cut}}$$=$500~keV, $i$$=$65\degr\ and $A_{\mathrm{Fe}}$$=$$1.0$ (see Section~\ref{subsec:refl}). The results of these fits are summarized in Table~\ref{tab:specchan}. We note that for these data the Fe abundance could be constrained when left free to vary. We obtained $A_{\mathrm{Fe}}$$=$$5.0\pm0.5$, which yielded an improved fit (with an $F$-test probability of $\simeq$$10^{-7}$; see Table~\ref{tab:specchan}). 

As can be seen in Table~\ref{tab:specchan}, the results of fitting the \chan\ and \nustar\ data together are similar to those obtained for the simultaneous \nustar/\swift\ data (cf. Table~\ref{tab:spec}). The excess above 5 keV in the \chan\ data is accounted for by including relativistic reflection modeled as \relx, although residual structure remains near 1--2 keV (see Figure~\ref{fig:specchan}, bottom). We also tried fits with \textsc{relconv*reflionx} as our relativistic reflection model, where we again fixed $i$$=$65\degr, $q$$=$3, $R_{\mathrm{out}}$$=$$500~\risco$, $a$$=$0.3, and $E_{\mathrm{cut}}$$=$500~keV, but left $A_{\mathrm{Fe}}$ free to vary. The results are included in Table~\ref{tab:specchan}. This reflection model seems to account better for the structure near 1~keV in the \chan\ data (Figure~\ref{fig:specchan}, bottom; see also Section~\ref{subsec:highres}). We note that the factor $\simeq$2 flux difference between the \chan\ and \nustar\ epochs could potentially result in differences in the reflection spectrum and could be a reason that some emission structure is not completely modeled.

 \begin{figure}
 \begin{center}
	\includegraphics[width=8.5cm]{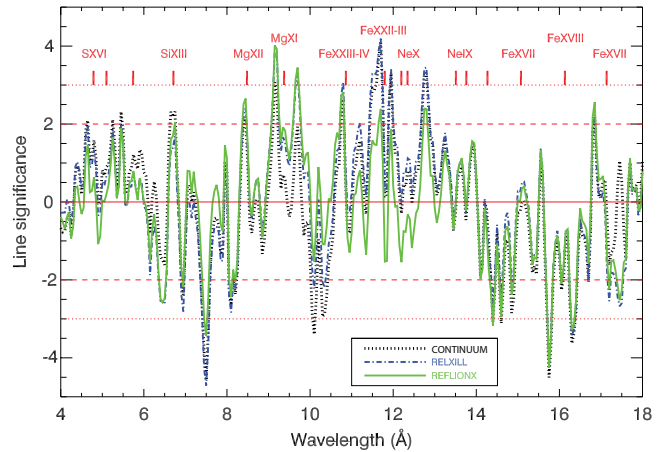}
    \end{center}
    \caption[]{Significance of line features in the \chan\ MEG and HEG data for different continuum models (see text for details). Positive and negative values indicate emission and absorption features, respectively. The rest wavelengths of abundant elements are indicated. The red dashed and dotted lines highlight 2$\sigma$ and 3$\sigma$ confidence levels, respectively. 
       }
 \label{fig:sign}
\end{figure}

%%%%

\subsection{\chan\ high-resolution X-ray spectrum}\label{subsec:highres}

\subsubsection{Line search}\label{subsubsec:linesearch}
We started our high-resolution spectral analysis with a phenomenological line search such as presented in \citet{pinto2016}. To this end we fitted the MEG and HEG data to an absorbed power law and black body continuum model, and added Gaussian lines with a fixed width of $1\,000~\kms$ over the 7--27~\AA\ range with increments of 0.05~\AA\ . The significance of the lines was computed by dividing the normalization by the 1$\sigma$ error. The results of our search are shown in Figure~\ref{fig:sign} as the black dotted line. This simple continuum description leaves several absorption (e.g. near 7.5 and 16~\AA; $\simeq$1.65 and 0.77~keV) and emission (e.g. near 12 and 13~\AA; $\simeq$1.03 and 0.95~keV) features that appear to be significant at a $>$3$\sigma$ level (see also the heavily binned spectral residuals in Figure~\ref{fig:lowElines}). 

Since the continuum modeling may affect the significance of any narrow spectral features, we repeated our line search after replacing the power-law component by \relx. For this reflection model we fixed all parameters except the normalization to the values obtained from fitting the \chan\ and \nustar\ data together with $A_{\mathrm{Fe}}$ free (Table~\ref{tab:specchan}). The result of this line search is shown as the blue dashed-dotted line in Figure~\ref{fig:sign}. We mostly pick up the same emission and absorption features as for the continuum modeling, of which some (near 12~\AA) appear more significant. 

Finally, we ran our line-search routine after using \textsc{relconv*relfionx}, rather than \relx, as the reflection component. All parameters except the normalization were fixed to the values listed in Table~\ref{tab:specchan}. This reflection model seems to account for the narrow lines near near 12 and 13~\AA\ ($\simeq$1.03 and 0.95~keV), as is illustrated by the green curve in Figure~\ref{fig:sign}. However, some of the other emission features remain also for this reflection model, e.g. near 9~\AA\ ($\simeq$1.38~keV). Using \textsc{relfionx} also reduces the significance of the absorption line near 7.5~\AA\ ($\simeq$1.65~keV). The only significant absorption features still present when using \textsc{relfionx} are lines near 16~\AA\ ($\simeq$0.77~keV), which have similar significance for the three different spectral models that we tried. We note, however, that the signal-to-noise ratio is low at $\gtrsim$15~\AA\ ($\lesssim$0.8~keV) and the continuum is less well constrained. 

The significances calculated in this analysis implicitly assume that the lines are at rest. If the lines are blue/redshifted this adds to the number of trials and hence decreases the significance. This may particularly affect the significance of the absorption lines, which appear to require a significant blueshift to correspond to any abundant element (see Section~\ref{subsubsec:highres}). Most of the emission lines, on the other hand, could be consistent with near-rest frame wavelengths of abundant elements, hence the significances calculated here should be a good approximation. Running our line-detection search for line widths of $250~\kms$ and $5\,000~\kms$, yielded a similar picture as that presented above for $1\,000~\kms$. 

Comparing the location of the various emission/absorption lines to the effective area of the MEG and HEG shows that none of the prominent features correspond to instrumental edges. Furthermore, we ran our line-search algorithm on \chan\ gratings data of the blazars H 1821+643 (obsIDs 1599, 2186, 2310), PKS 2155-304 (ObsIDs 337, 3167, 9712), and Mrk 421 (obsIDs 10663, 4148, 4149), which provided similar statistics as our data set and are not expected to show narrow spectral features. Indeed, we did not find any lines of comparable magnitude. Therefore, there are no obvious indications that the narrow features picked up by our line search are instrumental or due to statistical fluctuations.

 \begin{figure}
 \begin{center}
	\includegraphics[width=8.5cm]{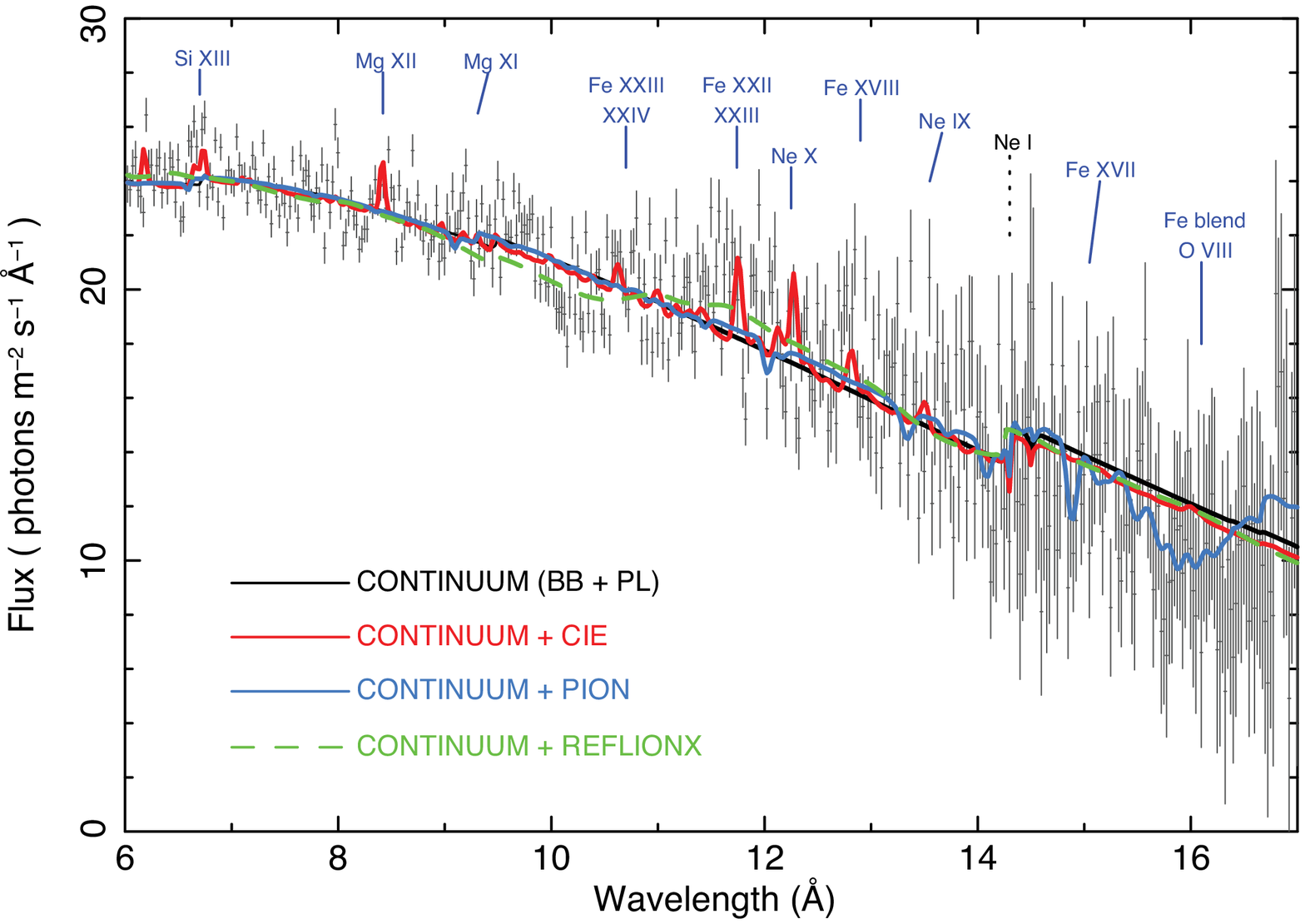}
    \end{center}
    \caption[]{Zoom of the \chan/MEG data fitted to different spectral models (see text for details). The position of several rest-frame transitions commonly seen in LMXBs (as well as active galactic nuclei) are indicated.
        }
 \label{fig:spechighres}
\end{figure}

%%%%

\subsubsection{Spectral modeling}\label{subsubsec:highres}
Assuming that the discrete emission and absorption lines found in our simple line search are real, we explored spectral models to probe what physical processes could cause these features. Unfortunately, due to the low flux and low significance of the lines, we cannot present a definitive solution. We explore different possibilities below and discuss some physical implications in Section~\ref{sec:discuss}. Since the fits are intended to be exploratory and are non-unique, we do not give errors on the fit parameters. In Figure~\ref{fig:spechighres} we show a zoom of the MEG data that captures our results.  

The black curve in Figure~\ref{fig:spechighres} shows a fit to a simple continuum consisting of a black body and a power law, which yields a $C$-stat value of 1447 for 1211 dof. As shown in Section~\ref{subsubsec:linesearch}, such a model would suggest the presence of some absorption and emission features. To explore the origin of the emission features, we added a \textsc{cie} component in our \textsc{Spex} fits, which describes the spectrum of a plasma in collisional-ionisation equilibrium. We adopted Solar abundances from \citet{lodders2009}. The \textsc{cie} model has three free parameters: the plasma temperature $kT_{\mathrm{cie}}$, the line broadening $\sigma_{\mathrm{cie}}$, and a normalization. We obtain $kT_{\mathrm{cie}}$$\simeq$1~keV, $\sigma_{\mathrm{cie}}$$\simeq$500--1000~$\kms$, and $C$-stat$=$ 1402 for 1208 d.o.f. This fit is shown as the red curve in Figure~\ref{fig:spechighres}. A collisional-ionized plasma would be able to account for several narrow emission lines.

To model the absorption features, we tried adding a photo-ionized absorber, \textsc{pion}, to an absorbed power law plus black body continuum. The \textsc{pion} model has four fit parameters: the ionization parameter $\log \xi$, the line broadening $\sigma_{\mathrm{pion}}$, the line-of-sight velocity $v_{\mathrm{pion}}$, and a normalization. We again adopted Solar abundances \citep[][]{lodders2009}. We obtain $\log \xi$$\simeq$2, $\sigma_{\mathrm{pion}}$$\simeq$500--1000~$\kms$, $v_{\mathrm{pion}}$$\simeq$$-(2000-3500)~\kms$ ($\simeq$0.01$c$), and $C$-stat$=$ 1422 for 1207 d.o.f. The best candidates for the 15.75~\AA\ and weaker 16.25~\AA\ absorption features are ``hot'' O\textsc{viii} or ``cooler'' Fe\textsc{xvi} (or lower Fe ions). The line broadening is similar to that obtained for the \textsc{cie} plasma model, which suggests that the emission and absorption lines could possibly arise from the same physical location. This fit is shown as the blue line in Figure~\ref{fig:spechighres}. Including photo-ionized absorption reproduces a relatively broad absorption feature near 16~\AA\ ($\simeq$0.77~keV) that was also found in our line search (Figure~\ref{fig:sign}; see also Figure~\ref{fig:lowElines}). The obtained line-of-sight velocity is negative, which would point to outflowing gas. 

The green dashed curve in Figure~\ref{fig:spechighres} shows a fit that includes reflection modeled as \textsc{reflionx}, which yields $C$-stat$=$1417 for 1207 dof. As this raises the flux near 12~\AA\ ($\simeq$1~keV) it can account for part of the emission excess. Nevertheless, some narrower emission features may not be accounted for (Section~\ref{subsubsec:linesearch}). If real, this un-modelled structure could perhaps be due to reflection that requires a narrower core (i.e. less relativistic blurring; more distant reflection) than supplied by the input model, or a different emission mechanism. Adding a \textsc{cie} model component results in $\Delta C$-stat$=$--27, and may indeed indicate that there are emission features that are not fully modeled by a single reflection zone.

%%%%%%%%%%%%%%%%%
% DISCUSSION
%%%%%%%%%%%%%%%%%

\section{Discussion}\label{sec:discuss}
In this work we presented \nustar, \swift\ and \chan\ observations of the neutron star LMXB \source, which has been persistently accreting at $\lx$$\simeq$$10^{-3} \ledd$ since 2006. The aim of this study was to gain more insight into the morphology of low-level accretion flows, and the nature of neutron star LMXBs that are able to accrete at these low rates for several years.

The 0.5--79~keV \nustar/\swift\ spectrum of \source\ can be modeled as a combination of a $\Gamma$$\simeq$2 power-law component and a soft, thermal component that can be described by a $kT$$\simeq$0.5~keV black body. This continuum spectral shape is very similar to that inferred from \xmm\ data obtained for a few neutron star LMXBs at $\lx$$\simeq$$10^{-4}~\ledd$ \citep[e.g.][]{armas2011,armas2013,armas2013_2,degenaar2013_xtej1709}. The temperature of the black body component appears to be too high for the measured X-ray luminosity, and the inferred radius to small, to be from the accretion disk. The thermal emission is therefore more likely coming from (part of) the neutron star surface, which is expected to become visible when moving to low $\lx$. We do not seem to detect the hard emission tail that is thought to be associated with surface accretion \citep[e.g.][]{deufel2001,dangelo2014,wijnands2015}. Possibly, at $\lx$$\simeq$$10^{-3}~\ledd$ this component is difficult to disentangle from the emission of the accretion flow itself.

The spectral data of \source\ show a broadened Fe-K line at $\simeq$6--7~keV. Such a feature has never been detected for a neutron star LMXB accreting at $\lx$$\lesssim$$10^{-2} \ledd$ before \citep[despite the availability of high-quality data; e.g.][]{armas2013,lotti2016}, but it is commonly seen in brighter LMXBs. Assuming that the line is due to relativistic disk reflection, we modeled the spectral data with \relx. This suggests that the inner disk radius was truncated away from the ISCO, at $R_{\mathrm{in}}$$\gtrsim$$20~\risco$ ($\gtrsim$$100~\gmc$ or $\gtrsim$225~km; 1$\sigma$ confidence). Due to the low flux of \source, however, the data quality is not good enough to rule out a location at the ISCO at $>$3$\sigma$ significance. Nevertheless, it would not be surprising to find a truncated disk at $\lx$$\lesssim$$10^{-2} \ledd$ and it is therefore interesting to further explore this (Section~\ref{subsec:innerdisk}). 

Our high-resolution \chan\ gratings data of \source\ reveal hints of discrete emission and absorption lines in the $\simeq$0.5--1.5~keV energy range. This includes what appears to be a collection of narrow emission lines near 12~\AA\ ($\simeq$1~keV) that can be modelled as disk reflection or a collisionally-ionized plasma. Unfortunately the significance of the narrow spectral features is low ($\lesssim$4.5$\sigma$) and depends on the underlying continuum/reflection spectrum. However, since this is the first gratings data of a neutron star LMXB accreting as low as $\lx$$\simeq$$10^{-3} \ledd$, it is interesting to explore plausible physical scenarios that could account for these lines (Section~\ref{subsec:outflow} and~\ref{subsec:emission}). The strongest and most robust feature found in our simple line search is a relatively broad feature near 16~\AA\ ($\simeq$0.77~keV), which has a significance of $\simeq$4$\sigma$ for all  spectral models that we explored (not accounting for trials). If the absorption is real, it can can be modelled as an outflowing photo-ionized plasma with a line-of-sight velocity of $\simeq$2\,000--3\,500~$\kms$ ($\simeq$0.01$c$). An outflow might be expected if the inner disk in \source\ is truncated due to the formation of a radiatively-inefficient accretion flow \citep[e.g.][]{narayan1994,blandford99,narayan2005}, or if the magnetosphere of the neutron star is acting as a propeller \citep[e.g.][]{illarionov1975,romanova2009,papitto2015_prop}.

%%%%

\subsection{Disk truncation at the magnetospheric boundary?}\label{subsec:innerdisk}
If the inner disk is indeed truncated at $R_{\mathrm{in}}$$\gtrsim$$100~\gmc$ in \source, this is a factor $\gtrsim$7 higher than typically found for non-pulsating neutron star LMXBs accreting at $\lx$$\gtrsim$$10^{-2}~\ledd$ \citep[$R_{\mathrm{in}}$$\simeq$5--15$~\gmc$; e.g.][]{cackett2010_iron,egron2013,miller2013_serx1,degenaar2015_4u1608,disalvo2015,ludlam2016,sleator2016}. It is also a factor $\gtrsim$3 higher than inferred for several (millisecond) X-ray pulsars that accrete at $\lx$$\gtrsim$$10^{-2}~\ledd$ \citep[$R_{\mathrm{in}}$$\simeq$15--30$~\gmc$; e.g.][]{miller2011,papitto2013_hete,king2016,pintore2016}. 

The black hole LMXB GX 339--4 shows a much narrower Fe-K line at $\simeq$$10^{-3} \ledd$ than at higher accretion luminosity. This can be interpreted as the inner disk receding from the ISCO to $R_{\mathrm{in}}$$\simeq$35~$\gmc$ as the mass-accretion rate drops, presumably due to disk evaporation \citep[][]{tomsick2009}. Our study suggests a larger truncation radius for \source\ at similar Eddington-scaled accretion rate. Disk evaporation should operate in neutron star LMXBs too, although it is expected to set in at lower $\lx$ than for black holes because soft photons emitted from the stellar surface cool the hot flow \citep[e.g.][]{narayan1995}. If the inner disk in \source\ is indeed further out than in GX 339--4, this might point to a different truncation mechanism.

We note that there is degeneracy in inferring a truncated disk from reflection modeling \citep[e.g.][]{fabian2014}. In particular, in case of GX 339--4 it has been pointed out that the narrowing of the Fe-K line with decreasing $\lx$ could also be due to the illuminating X-ray source moving away, i.e. an increasing height of the corona in a ``lamp-post'' geometry \citep[][]{dauser2013}. It is, however, not obvious that in neutron star LMXBs the accretion disk is also illuminated by a corona (rather than e.g. the boundary layer) and hence that a lamp-post geometry would apply for \source. 

In neutron star LMXBs, it is also possible that the stellar magnetic field truncates the inner accretion disk. In fact, a magnetically-inhibited accretion flow has been proposed as a possible explanation for the sustained low accretion rate of some VFXBs like \source\ \citep[][]{heinke09_vfxt,heinke2014_gc,degenaar2014_xmmsource}. If the blue-shifted absorption in the \chan\ data is real, this could possibly form a consistent physical picture in which the accretion flow is stopped at the magnetospheric boundary that acts as a propeller. This is an interesting scenario because the inferred inner disk radius would then provide constraints on the magnetic field strength and spin period of the neutron star.

%%%%

\subsection{Estimates of the neutron star magnetic field strength}\label{subsec:Bfield}
If the inner accretion disk in \source\ is truncated at the magnetospheric radius, we can estimate the magnetic field strength. To this end we use equation (1) from \citet{cackett2009_iron}, which is based on the derivations of \citet{ibragimov2009}, to write the following expression for the magnetic field strength:
\begin{eqnarray*}
B = 1.2\times10^{5} \, k_{\mathrm{A}}^{-7/4} \, \left(\frac{R_{\mathrm{in}}}{GM/c^2}\right)^{7/4} \, \left(\frac{M}{1.4~\mathrm{M_\odot}}\right)^2 \, \frac{D}{5~\mathrm{kpc}} \nonumber \\ 
\times \left(\frac{R}{10^6~\mathrm{cm}}\right)^{-3} \, \left( \frac{f_{\mathrm{ang}}}{\eta} \frac{F_{\mathrm{bol}}}{10^{-9}~\mathrm{erg~cm^{-2}~s^{-1}}} \right)^{1/2}~\mathrm{G},
\end{eqnarray*}

\noindent
where $f_{\mathrm{ang}}$ is an anisotropy correction factor \citep[which is close to unity;][]{ibragimov2009}, $k_{\mathrm{A}}$ a geometry coefficient \citep[expected to be $\simeq$0.5--1.1;][]{psaltis1999,long2005,kluzniak2007}, and $\eta$ the accretion efficiency. 

We use $D$$=$5~kpc, $M$$=$1.4$~\Msun$, $R$$=$10~km, $R_{\mathrm{in}}$$\gtrsim$100~$\gmc$, and conservatively assume that the bolometric flux is equal to the 0.5--79 keV flux determined from our joint \nustar/\swift\ fits (i.e. $F_{\mathrm{bol}}$$=$$F_{\mathrm{0.5-79}}$$\simeq$$1.2\times10^{-10}~\flux$). Furthermore, we assume $f_{\mathrm{ang}}$$=$1, $k_{\mathrm{A}}$$=$1, and $\eta$$=$0.1. We then obtain $B$$\gtrsim$$4\times10^{8}$~G for \source. This is a factor of a few higher than typical estimates for neutron stars in LMXBs, although within the maximum allowable range determined in a recent analysis of the coherent timing properties of several millisecond X-ray pulsars \citep[][]{mukherjee2015}.

%%%%

\subsection{A propeller-driven outflow?}\label{subsec:outflow}
If the accretion disk in \source\ is indeed truncated {\it and} this is due to the magnetic field of the neutron star, the rotating magnetosphere may act as a propeller \citep[e.g.][]{illarionov1975,lovelace1999,romanova2009,papitto2015_prop}. Magnetohydrodynamic simulations show that an active propellor can cause a two-component outflow consisting of an axial jet and a conical wind \citep[e.g.][]{romanova2009}. The wind component has a high density, outflow velocity of $\simeq$0.03$c$--0.1$c$, and is shaped like a thin conical shell with a half-opening angle of $\simeq$30$\degr$--40$\degr$. The jet component has a lower density and a higher outflow velocity ($\simeq$0.4$c$--0.6$c$). 

If the blue-shifted absorption in our \chan\ data is real, a line-of-sight velocity of $\simeq$0.01$c$ could potentially be consistent with a wind driven by an active propeller. Interestingly, a small subgroup of neutron star LMXBs that appear to exhibit propeller stages, the transitional millisecond radio pulsars \citep[e.g.][]{archibald2009,papitto2014,papitto2015_prop}, seem to have more luminous radio jets than other neutron star LMXBs \citep[][]{deller2014}. If a propeller is operating in \source, it may thus be expected to exhibit a strong radio jet too.

We note that even if the magnetic field is truncating the inner accretion disk, it is not necessary that a propeller is operating. Another possibility is a ``trapped disk'' morphology \citep[e.g.][]{dangelo2010,dangelo2012}. In the propellor scenario, strong outflows are formed and little matter accretes on to the neutron star so that the accretion flow may dominate the overall X-ray luminosity. For a trapped disk, however, only a weak outflow is expected and considerable amounts of gas can still accrete on to the neutron star magnetic poles, which may dominate the overall X-ray luminosity.

%%%%

\subsection{Estimates of the neutron star spin period}\label{subsec:spin}
If the blue-shifted absorption in our \chan\ data is real {\it and} due to a propeller-driven outflow, the assumption that the inner accretion disk is truncated at the magnetospheric boundary allows to put some constraints on the neutron star spin period. A neutron star is thought to be in the propellor regime when the magnetospheric radius is larger than the co-rotation radius. At this radius the Keplerian orbital velocity of the matter equals the rotational velocity of the neutron star, i.e. $R_\mathrm{co}$$=$$(GMP_{\mathrm{s}}^2/4\pi^2)^{1/3}$, where $P_{\mathrm{s}}$ is the spin period of the neutron star. Assuming that the inner disk radius is truncated by the magnetosphere, i.e. $R_\mathrm{m}$$=$$R_\mathrm{in}$$\gtrsim$$100~\gmc$ ($\simeq$225~km), the requirement that $R_\mathrm{m}$$>$$R_\mathrm{co}$ suggests that \source\ is in the propellor regime if $P_{\mathrm{s}}$$\lesssim$19~ms. 

We seem to detect thermal emission from the stellar surface in our X-ray spectra and two thermonuclear X-ray bursts have been detected from \source\ \citep[][]{degenaar2013_igrj1706,iwakiri2015,negoro2015_igrj1706}. This implies that at least some matter must be able to accrete on to the neutron star. The accretion disk must therefore lie within the light cylinder radius, where the rotational velocity of the magnetic field lines reaches the speed of light, i.e. $R_\mathrm{lc}$$=$$c/\Omega$ with $\Omega$$=$$2\pi/P_{\mathrm{s}}$ being the angular velocity. If $R_\mathrm{in}$$\gtrsim$$100~\gmc$ ($\simeq$225~km), this would suggest that $P_{\mathrm{s}}$$\gtrsim$4.7~ms. 

If the accretion disk is truncated by the magnetic field and a propellor operates, that would require $P_{\mathrm{s}}$$\simeq$4.7--19~ms for \source. This is within the typical range of spin periods measured for neutron stars LMXBs from coherent X-ray pulsations or burst oscillations \citep[1.6--10~ms; e.g.][for a list]{patruno2010}. 

%%%%

\subsection{Low-energy narrow X-ray emission lines}\label{subsec:emission}
Binning the \chan/HETG data reveals an emission feature near 1~keV. A broad line around the same energy was detected in the \swift/XRT data obtained during the energetic X-ray burst of \source\ in 2012 \citep[$E_{\mathrm{l}}$$=$$1.018\pm 0.004$~keV, and EW$=105 \pm 3$~eV;][]{degenaar2013_igrj1706}. It was interpreted as Fe-L or Ne \textsc{x} emission arising from irradiation of relatively cold gas orbiting at a distance of $\simeq$$10^{3}$~km ($\simeq$$500~\gmc$) from the neutron star (by assuming that the line was rotationally-broadened by gas moving in Keplerian orbits). Broad emission lines near 1 keV have been detected in the accretion spectra of a number of other neutron star LMXBs \citep[e.g.][]{vrtilek1991,kuulkers1997,diaztrigo2006,cackett2010_iron,papitto2013_hete}. 

Exploiting the high spectral resolution of the HETG, we found that in \source\ the emission feature near 1~keV ($\simeq$12~\AA) may be resolved into a number of narrow lines. Narrow emission lines also appear to be present at other energies (e.g. near 9 and 10~\AA; $\simeq$1.38 and 1.24~keV). High-resolution observations of some other neutron star LMXBs revealed complexes of narrow emission lines at low energies, typically consistent with being at rest \citep[e.g.][]{cottam2001_exo,cottam2001,schulz2001,beri2015}. Proposed explanations include a pulsar-driven disk wind or photo-ionized emission from a thickened structure in the accretion disk (e.g. the impact point where the gas stream from the companion hits the outer accretion disk). 

In case of \source, the strongest narrow emission line is located at $\simeq$11.6~\AA\ ($\simeq$1.07~keV). If real, it could correspond to Fe-L at rest. This would render a collisionally-ionized plasma more likely; photo-ionized gas lines from lower-Z elements (e.g. O, Ne) should be stronger than Fe-L, which doesn't seem to be the case for our data. Perhaps shocks resulting from the accretion flow running into the magnetosphere or from matter impacting the magnetic poles could give rise to collisionally-ionized emission in this neutron star LMXB. Alternatively, this line could correspond to Ne\,{\sc x} blue-shifted by $\simeq$0.045$c$ ($\simeq$$13.5\times10^{3}~\kms$), which would be indicative of an outflow. A third, perhaps more likely, possibility is that the emission lines are due to reflection. However, a single reflection component that also fits the Fe-K line seems to leave excess emission near 1~keV. This could indicate that there are multiple reflection zones, or that different emission mechanisms are responsible for the different lines.

%%%%

\section*{Acknowledgements}
ND is supported by an NWO Vidi grant and an EU Marie Curie Intra-European fellowship (contract no. FP-PEOPLE-2013-IEF-627148). ND acknowledges valuable discussions with Anne Lohfink, Victor Doroshenko, and Caroline D'Angelo. CP and ACF are supported by ERC Advanced Grant Feedback 340442. JMM acknowledges support from the \chan\ guest observer program. DA acknowledges support from the Royal Society. RW is supported by an NWO Top grant, module 1. We thank the referee for useful comments. This work is based on data from the \nustar\ mission, a project led by California Institute of Technology, managed by the Jet Propulsion Laboratory, and funded by NASA. We thank Neil Gehrels and the \swift\ duty scientists for rapid scheduling of observations and acknowledge the use of the Swift public data archive. 

\footnotesize{

}

\end{document}